\documentclass[12pt]{article}
\usepackage{graphicx}
\graphicspath{ {images/} }

\begin{document}
\setcounter{page}{1} \pagestyle{plain} \vspace{1cm}
\begin{center}
\Large{\bf A Note on Coherent States with Quantum Gravity Effects}\\
\small \vspace{1cm}
{\bf N. Sadeghnezhad\footnote{Email: nsadegh@riaam.ac.ir}}\quad \\
\vspace{0.5cm} {\it Research Institute for Astronomy and
Astrophysics of Maragha (RIAAM)- Maragha,\\
IRAN, P. O. Box: 55134-441}

\end{center}
\vspace{1.5cm}
\begin{abstract}
Existence of a minimal measurable length and an upper bound for the
momentum fluctuations are the casting reasons for generalization of
uncertainty principle and then reformulation of Hilbert space
representation of quantum mechanics. In this paper, we study the
consequences of the Generalized Uncertainty Principle (GUP) in the
presence of both minimal length and maximal momentum. We consider a
simple harmonic oscillator in the framework of GUP by introducing
it's energy eigenstates and energy spectrum. Investigation of
coherent states for a generalized harmonic oscillator and it's
generic behavior are the other topics in our study.
\end{abstract}
\vspace{1.5cm}
\newpage

\section{Introduction}
\quad It is a well-known idea that gravity may modify the
uncertainty principle. Bearing the gravity in mind, field theory
leads to an effective cutoff (minimum measurable length) in the
ultraviolet energies. In fact, since high energies have major
gravitational effects, they will change the spacetime structure in
the small scales [1]. Theoretically, the minimum measurable length
is proposed in various approaches to the quantum gravity such as
string theory [2-6], loop quantum gravity [7] and quantum geometry
[8], which has the same order as the Planck length ($l_{pl}\sim
10^{-35}\textmd{m}$) [9-11]. In the Heisenberg uncertainty principle
(HUP), one cannot measure the momentum and position of a particulary
particle with zero uncertainty, together. However, it is possible to
measure one of these observable quantities by withdrawing the
corresponding information about the other. It means that one can
vanish the uncertainty in position yielding $(\Delta x)_{min}=0$.
But the story is changed by considering the modified uncertainty
principle due to the quantum gravity which is now called the
Generalized Uncertainty Principle (GUP). In fact, due to the
non-zero effective cutoff, it is not possible to set $(\Delta
x)_{min}=0$ whenever the effects of quantum gravity become important
meaning that we should replace HUP by GUP [12-26]. Moreover,
Hamiltonian is changed due to GUP leading to the changes of the
energy spectrum of quantum systems. Since GUP implies $(\Delta
x)_{min}\neq0$, one may conclude that spacetime has a
non-commutative structure in the Planck scales leading to quantize
the spacetime in the Planck scales. Therefore, one can reinterpret
this cutoff as the quanta of space, which is due to the quantum
fluctuations of background spacetime, leading to a new
representation for the Hilbert space [27]. This new representation
attracted more investigators to itself [28-30]. Moreover,
considering the Planck length as the minimum permissible length,
which is independent of the observer, changes the Special Relativity
foundation that is yielding the Doubly Special Relativity (DSR)
theory [31-34]. In the DSR theory, the minimum effective cutoff
makes an upper bound on momentum. Therefore, the Planck length
inspires an upper bound for the admissible energy in this theory. We
should note that the above results are modified to the more common
situations in which the spacetime curvature is taken into account.
Indeed, the spacetime curvature induces a non-zero minimum to the
momentum uncertainty. The non-zero uncertainty in the position and
momentum lead to retire the wave functions in the position and
momentum spaces, which are introduced in the quantum mechanics based
on HUP, respectively. In order to avoid these shortcomings, bearing
the states with maximum localization in mind, one should use the
quasi-position and quasi-momentum representations and reformulate
the quantum mechanics, and therefore the Hilbert space [27,35]. Our
aim in this paper is investigating the effects of considering the
minimal length and maximal momentum on the harmonic oscillator and
its properties. The paper is organized as follows. In the next
section, we review GUP by considering the minimum length and maximum
momentum considerations. In section 3 we present the generalized
harmonic oscillator as well as its eigenstates and eigenvalues.
Coherent states of a harmonic oscillator is presented in section 4.
In addition, we study the normalization coefficient and probability
distribution of the coherent states. The last section is devoted to
a summary and concluding remarks.

\section{GUP with minimal length  and maximal momentum}
\quad In the framework of a generalized uncertainty principle that
predicts maximal observable momentum in addition to minimal
observable length, we can write [19-24,35]
\begin{equation}
\Delta x \,\ \Delta p \geq \frac{ \hbar}{2} (1- 2\alpha <p>
+4\alpha ^{2} < p^{2} >) \, ,
\end{equation}
which $ \alpha $ is the GUP parameter in the presence of the two
aforesaid cutoffs. The above uncertainty relation can be obtained
from the following algebraic structure
\begin{equation}
[x,p]= i \hbar (1- \alpha p +2 \alpha ^{2} p^{2}).
\end{equation}
In this relation there is a first order term in particle's momentum
which has its origin on the existence of a maximal momentum, whereas
the second order term in particle's momentum originates from the
existence of a minimal length.
\\ We can define position and momentum operators for the GUP case as
\begin{equation}
X=x \quad\quad  , \quad\quad
P = p(1- \alpha p +2 \alpha ^{2} p^{2}),
\end{equation}
where  $x$ and $p$ ensure the Jacobi identities, and $X$ and $P$
satisfy the generalized commutation relation
\begin{equation}
[X,P]= i \hbar (1- \alpha p +2 \alpha ^{2} p^{2}).
\end{equation}
In this case, we interpret $p$ as the momentum operator at low
energies by a standard representation in position space,
$p_{j}=\frac{\hbar}{i}\frac{\partial}{\partial x_{j}}$, and $P$ as
the momentum operator at high energies which has the generalized
representation in position space as
$$P_{j}=\frac{\hbar}{i}\frac{\partial}{\partial x_{j}}[1-\alpha(\frac{\hbar}{i}\frac{\partial}{\partial x_{j}})+2\alpha^{2}(\frac{\hbar}{i}\frac{\partial}{\partial x_{j}})^{2}] \, .$$
To show how maximal momentum arises in this setup, we first find the
minimal observable length, i.e. $ \Delta x_{0}\equiv\Delta x
_{min}(<p>=0)$. We can write the inequality (1) on the boundary of
the allowed region and use $(\Delta p)^{2}=<p^{2}>-<p>^{2}$ to
obtain a second order equation for $\Delta p$, which has the
following solutions for $\Delta p$
\begin{equation}
\Delta p=\frac{\Delta x}{4\alpha ^{2} \hbar}\pm\sqrt{(\frac{\Delta x}{4\alpha ^{2} \hbar})^{2}-\frac{<p>}{2\alpha}(2\alpha <p>-1)-\frac{1}{4\alpha ^{2}}} \: .
\end{equation}
The reality of solutions gives the minimum value of $\Delta x$ as
$$\Delta x_{min}(<p>)=2\alpha \hbar \sqrt{1-2\alpha <p>+4\alpha ^{2}<p>^{2}} \; . $$
Using $<p>=0$, absolutely smallest uncertainty in position (absolute
minimal observable length) can be deduced from the latter equation
as
\begin{equation}
\Delta x_{0}=2\alpha\hbar \, .
\end{equation}
Due to duality of position and momentum operators, we can assume
$\Delta x_{min} \propto \Delta p_{max}$. With this assumption, using
the condition $<p>=0$ in Eq. (5) and making use of Eq. (6), we have
\begin{equation}
(\Delta p)_{max}=\frac{1}{2\alpha} \, ,
\end{equation}
where we will assume this result as the maximal measurable momentum
in our setup.

\subsection {Representation on momentum space}
\quad It is to be noted that, if we assume the minimal observable
length as minimal, nonzero uncertainty in position, we have no
longer a Hilbert space representation on position space wave
functions of the ordinary quantum mechanics. This is because one can
not find any physical state which is a position eigenstate, since
such an eigenstate would have zero uncertainty in position.
Therefore we must construct a new Hilbert space representation
compatible with relation (4). This representation can be achieved in
a continuous momentum space. Now, in this space, momentum and
position operators have the form
\begin{equation}
P=p \quad\quad  , \quad\quad  X = (1- \alpha p +2 \alpha ^{2} p^{2})x \, ,
\end{equation}
where $x=i \hbar\frac{\partial}{\partial p}$. Due to the presence of
the additional factor $(1-\alpha p +2\alpha ^{2} p^{2})$, the scalar
product in momentum representation should be rewritten as
\begin{equation}
 < \phi \mid \varphi  > =\int^{+P_{pl}} _{-P_{pl}} \frac{ dp}{1- \alpha p + 2 \alpha
 ^{2} p^{2}}\phi^{*}(p) \varphi(p) \, .
\end{equation}
We note that appearance of the limits  $- P_{pl}$ to $+ P_{pl}$
(Planck momentum) originate from the maximal measurable momentum, so
that in the absence of this cutoff the integrals must be calculated
from $- \infty$ to $ + \infty$ [27]. In present framework, the
identity operator would be modified as
\begin{equation}
1 =\int^{+P_{pl}} _{-P_{pl}} \frac{ dp}{1-  \alpha p
+2 \alpha ^{2} p^{2}} \mid p><p \mid.
\end{equation}

\subsection {Functional analysis of the position operator }
\quad The position operator ( X in Eq. (8) ) acting on position
eigenstate in momentum space, $ \varphi _{ \xi} (p) = <p \mid \xi >
$, gives the following eigenvalue equation
\begin{equation}
i \hbar(1- \alpha p + 2 \alpha ^{2} p^{2}) \frac{
 \partial \varphi_{ \xi}(p)}{ \partial p}= \xi \varphi_{ \xi}(p) \, .
\end{equation}
By solving this differential equation, we obtain the position
eigenvectors in the presence of aforesaid cutoffs as
\begin{equation}
\varphi _{ \xi} (p)= \varphi _{\xi} (0) \exp \Bigg[-i \frac{2
\xi}{\alpha \hbar \sqrt{7}} \Bigl\{ { \tan^{-1} (\frac{ 1 }{ \sqrt{7}})} +  \tan^{-1}  (\frac{ 4 \alpha p -1}{
\sqrt{7} }) \Bigr\} \Bigg].
\end{equation}
Using the normalization
$$
1= < \varphi \mid \varphi> = \int^{+P_{pl}} _{-P_{pl}} \frac{
1}{1- \alpha p + 2 \alpha
 ^{2} p^{2}}\phi^{*}_{ \xi}(p) \phi_{ \xi}(p) dp \, ,
$$
we can obtain the coefficients $\varphi_{\xi}(0)$ as
$$
 \varphi_{\xi}(0) =\sqrt{\frac{\alpha \sqrt{7}
}{ 2}} \Bigg[ { \tan^{-1} (\frac{ 4 \alpha P_{pl} -1}{
\sqrt{7}})} +  \tan^{-1}  (\frac{ 4 \alpha P_{pl} +1}{ \sqrt{7} })
\Bigg]^{-1/2}.
$$

\subsection {Maximal localization }
\quad In the presence of minimum observable length, $l_{pl}=\Delta
x_{0}=2\alpha \hbar$, it is not possible to probe distances less
than Planck length. So, the notion of spacetime manifold should be
revised for the finite resolution of the spacetime points. In this
manner, we are obliged to introduce the states with maximal
localization that are confined up to Planck length and it is
impossible to localize them further. Now, we consider the states
$\mid\varphi^{ml} _{ \xi}>$\ of maximal localization around a
position $\xi \geq l_{pl}$ and write
\begin{equation}
<X>=< \varphi^{ml} _{ \xi} \mid X \mid \varphi^{ml} _{ \xi} > = \xi .
\end{equation}
By using the positivity of norm for each state $\mid \varphi >$ in
the representation of the Heisenberg algebra, namely
$$
\parallel \Big( X- <X> + \frac{< [X,P]>}{2 ( \Delta P)^{2}}( P- <P>) \Big) | \varphi>
\parallel \geq 0 ,
$$
we can deduce [27] that the state $\mid \varphi >$ will be on the
boundary of the physically allowed region only it obeys
\begin{equation}
\Big( X- <X> + \frac{< [X,P]>}{2 ( \Delta P)^{2}}( P- <P>) \Big) | \varphi> = 0 .
\end{equation}
Using Eq. (8), relation (14) takes the form of a differential
equation in momentum space as
\begin{equation}
\Big(  i \hbar (1- \alpha p + 2 \alpha
 ^{2} p^{2})\frac{ \partial}{ \partial p}- <X> +
 i \hbar \frac{1+ 2 \alpha^{2} (\Delta p)^{2} + 2 \alpha ^{2} <p>^{2} - \alpha <p>}{2 ( \Delta p)^{2}}( p- <p>) \Big)| \varphi>
 = 0.
\end{equation}
By taking into account that the states of absolutely maximal
localization can only be obtained for $<p>=0$, and using Eqs. (7)
and (13), Eq. (15) can be solved to obtain
$$
\varphi_{ \xi}^{ml}(p)=\varphi_{ \xi}^{ml}(0)\frac{
e^{-\frac{3}{2\sqrt{7}}(\frac{4i \xi}{ 3 \alpha \hbar}+1)(\tan^{-1} (\frac{1}{\sqrt{7}})+\tan^{-1}
(\frac{4 \alpha p-1}{ \sqrt{7}}))}}{ (1- \alpha p + 2 \alpha
 ^{2} p^{2} )^{ \frac{3}{4}}} .
$$
Using normalization condition
$$
1= < \varphi_{ \xi}^{ml}(p) |\varphi_{ \xi}^{ml}(p)>=
\int_{-P_{pl}}^{+P_{pl}} \varphi_{ \xi} ^{ml}(0) \varphi_{ \xi}
^{ml^{*}}(0) \frac{e^{-\frac{3}{\sqrt{7}}(\tan^{-1} (\frac{1}{\sqrt{7}})+\tan^{-1}
(\frac{4 \alpha p-1}{ \sqrt{7}}))}}{(1- \alpha p + 2 \alpha
 ^{2} p^{2})^{\frac{5}{2}}} dp ,
$$
we find $\varphi ^{ml}_{\xi}(0)$ as
$$
\phi ^{ml}_{\xi}(0) =\sqrt{6 \alpha} \Big[ \sqrt{8} e^{\eta \tan^{-1} (\eta)} - e^{-\eta \tan^{-1}(\frac{\eta}{3})}\Big]^{- \frac{1}{2}} e^{\frac{\eta}{2} \tan^{-1}(\frac{\eta}{3})} ,
$$
where $\eta\equiv\frac{4\alpha P_{pl}-1}{\sqrt{7}}$. Since we have $
P_{pl}=\frac{1}{2\alpha}$, we find $ \eta=\frac{3}{\sqrt{7}}$.
Finally, the momentum space wavefunction for maximally localized
states around $\xi$ can be written in the form
\begin{equation}
 \phi ^{ml}_{\xi}(p) =\frac{ \sqrt{6 \alpha} \Big[ \sqrt{8} e^{\eta \tan^{-1}(\eta)}-e^{-\eta \tan^{-1}(\frac{\eta}{3})} \Big]^{- \frac{1}{2}}}{(1- \alpha p + 2 \alpha^{2} p^{2})^{\frac{3}{4}}}
e^{-\frac{\eta}{2} \tan^{-1}(\frac{4 \alpha p -1}{\sqrt{7}})} e^{-\frac{2i\xi}{\alpha \hbar \sqrt{7}}(\tan^{-1}( \frac{\eta}{3})+\tan^{-1}(\frac{4\alpha p-1}{\sqrt{7}}))}.
\end{equation}
In ordinary quantum mechanics, we expand the states $\mid \varphi >$
in the position eigenbasis $\{ \mid x > \}$ as $< x \mid \varphi >$.
But, there are now no physical states which would form a position
eigenbasis. However, there is a possibility to project arbitrary
states $\mid \varphi>$ on maximally localized states, $\mid \varphi
^{ml}_{\xi} >$, to obtain the probability amplitude of maximal
localization for the particle around the position $\xi$. We will
call the collection of these projections,
$<\varphi_{\xi}^{ml}|\phi>$, the states " quasi-position wave
function"
$$
\phi( \xi) := <\varphi_{\xi}^{ml}|\phi> .
$$
Using a generalization of the Fourier transformation that maps
momentum space wavefunction into quasi-position space wavefunction,
we can transform a state's wavefunction in momentum representation
to it's quasi-position wavefunction as follows
\begin{equation}
\varphi(\xi)=\varphi_{ \xi}^{ml}(0)\int _{-P_{pl}}^{+P_{pl}}\frac{e^{(-\frac{\eta}{2}+\frac{2i\xi}{\alpha\hbar\sqrt{7}})(\tan^{-1}(\frac{\eta}{3})+\tan^{-1}(\frac{4\alpha p-1}{\sqrt{7}}))}
}{(1- \alpha p + 2 \alpha
 ^{2} p^{2})^{\frac{7}{4}}}\varphi(p)dp.
\end{equation}
As in ordinary quantum mechanics, one can write
$$
e^{i \frac{2\xi}{\alpha\hbar\sqrt{7}} (tan^{-1}(\frac{\eta}{3})+tan^{-1}(\frac{4\alpha p-1}{\sqrt{7}}))}\equiv e^{iK\xi} ,
$$
to take into account the $K\equiv \frac{2}{\alpha \hbar \sqrt{7}} \Bigl( tan^{-1}(\frac{\eta}{3})+tan^{-1}(\frac{4\alpha p-1}{\sqrt{7}}) \Bigr)$ as modified wavenumber. So, the modified wavelength for quasi-position wavefunction of physical states has the form
$$
\lambda(p)=\frac{\pi \alpha \hbar \sqrt{7}}{\tan^{-1}(\frac{\eta}{3})+\tan^{-1}(\frac{4\alpha p -1}{\sqrt{7}})}.
$$
Since $\alpha \neq 0$ and $p$ is limited to the Planck momentum,
there is no wavelength smaller than $\lambda _{0}$
\begin{equation}
\lambda _{0}=\lambda(P_{pl})=\frac{\pi \alpha \hbar \sqrt{7}}{\tan^{-1}(\frac{\eta}{3})+\tan^{-1}(\frac{4\alpha P_{pl} -1}{\sqrt{7}})} .
\end{equation}
Using the relation between momentum and energy,
$E=\frac{p^{2}}{2m}$, we can write the maximum energy of the
momentum eigenstates as
\begin{equation}
E(\lambda_{0})=\frac{P_{pl}^{2}}{2m} ,
\end{equation}
which for $m\approx M_{pl}$, the energy of short wavelength modes
will be the Planck energy, $E(\lambda _{0})\approx E_{pl}$. Note
that there is not any energy divergency in $\lambda _{0}$. This
result is in agreement with ordinary quantum mechanics and is an
important outcome of a GUP formalism in the presence of both minimal
length and maximal momentum. By inverse Fourier transform of Eq.
(17), we have
\begin{equation}
\phi (p)= \Big (\varphi^{ml} _{\xi}(0)
^{-1}\Big)\int^{+\infty}_{-\infty}\frac{(1-
\alpha p +2\alpha ^{2}p^{2})^{\frac{3}{4}}}{2\pi\hbar} e^{(\frac{\eta}{2}-\frac{2i\xi}{\alpha\hbar\sqrt{7}})(\tan^{-1}(\frac{\eta}{3})+\tan^{-1}(\frac{4\alpha p-1}{\sqrt{7}}))} \phi(\xi)d\xi \, ,
\end{equation}
which because of the integration over $\xi$ (not p), the integration
interval will be over $-\infty$ to $+\infty$. Using Eq. (20) and
following the customary method in ordinary quantum mechanics, we can
deduce the generalized form of momentum operator in the
quasi-position space. Since
$$
\frac{\partial}{\partial\xi} e^{i\frac{2\xi}{\alpha\hbar \sqrt{7}}(\tan^{-1}(\frac{\eta}{3})+\tan^{-1}(\frac{4\alpha p-1}{\sqrt{7}})}=i\frac{2}{\alpha\hbar\sqrt{7}}\Big( \tan^{-1}(\frac{\eta}{3})+\tan^{-1}(\frac{4\alpha p-1}{\sqrt{7}} \Big) e^{i\frac{2\xi}{\alpha\hbar \sqrt{7}}(\tan^{-1}(\frac{\eta}{3})+\tan^{-1}(\frac{4\alpha p-1}{\sqrt{7}})} ,
$$
one can infer the relation $\frac{\alpha \sqrt{7}}{2} \frac{\hbar}{i} \frac{\partial}{\partial \xi}\equiv \tan^{-1}(\frac{\eta}{3})+\tan^{-1}(\frac{4\alpha p-1}{\sqrt{7}})$. Then, using $\tan(\tan^{-1}(\frac{\eta}{3})+\tan^{-1}(\frac{4\alpha p-1}{\sqrt{7}}))=\frac{\sqrt{7}\alpha p}{2-\alpha p}$ momentum operator can be obtained as
\begin{equation}
P\equiv\frac{2}{\alpha} \frac{\tan(\frac{\alpha \sqrt{7}}{2}\frac{\hbar}{i}\frac{\partial}{\partial\xi})}{\sqrt{7}+\tan(\frac{\alpha \sqrt{7}}{2}\frac{\hbar}{i}\frac{\partial}{\partial\xi})}.
\end{equation}

\section{ Generalized harmonic oscillator }
\quad In this section we generalize the formulation of a linear
harmonic oscillator in the presence of both minimal length and
maximal momentum, and obtain the eigenvalue and eigenfunctions of
harmonic oscillator by solving the Schr\"{o}dinger equation.
According to position and momentum operators in momentum space
representation ( Eq. (8) ) and by using them in harmonic oscillator
Hamiltonian, $ H=\frac{P^{2}}{2m}+\frac{1}{2}m \omega^{2} X^{2} $,
time-independent Schr\"{o}dinger equation, $H\psi =E\psi$, can be
written as
\begin{equation}
\frac{\partial ^{2} \psi (p)}{\partial p^{2}}+\frac{4\alpha ^{2} p-\alpha}{1-\alpha p+2\alpha ^{2} p^{2}}\frac{\partial \psi (p)}{\partial p}+\frac{1}{(1-\alpha p+2\alpha ^{2} p^{2})^{2}}(\epsilon -\beta ^{2} p^{2})\psi (p)=0 ,
\end{equation}
with
$$
\epsilon=\frac{2E}{m \omega ^{2} \hbar ^{2}} \quad\quad\quad   , \quad\quad\quad  \beta ^{2}=\frac{1}{(m\hbar \omega)^{2}} .
$$
With solving the above differential equation, the eigenfunctions can
be obtained in terms of the Legendre functions
$$
\psi(p)=C_{1}\exp \Bigg[ -\frac{1}{4}\frac{\sqrt{2}\tanh^{-1}\frac{\sqrt{2}\alpha p}{\sqrt{\alpha p-1}}}{\sqrt{\alpha p-1}} \Bigg] P \Bigg( \frac{1}{2}\frac{\sqrt{\alpha^{4}+\beta^{2}}-\alpha^{2}}{\alpha^{2}},\frac{1}{4}\frac{\sqrt{2}\sqrt{\alpha^{4}-4\epsilon\alpha^{2}+2\beta^{2}(\alpha p-1)}}{\alpha^{2}\sqrt{\alpha p-1}},\frac{2\alpha p}{\sqrt{2\alpha p-2}} \Bigg)
$$
\begin{equation}
\quad\quad +C_{2}\exp \Bigg[ -\frac{1}{4}\frac{\sqrt{2}\tanh^{-1}\frac{\sqrt{2}\alpha p}{\sqrt{\alpha p-1}}}{\sqrt{\alpha p-1}} \Bigg] Q \Bigg( \frac{1}{2}\frac{\sqrt{\alpha^{4}+\beta^{2}}-\alpha^{2}}{\alpha^{2}},\frac{1}{4}\frac{\sqrt{2}\sqrt{\alpha^{4}-4\epsilon\alpha^{2}+2\beta^{2}(\alpha p-1)}}{\alpha^{2}\sqrt{\alpha p-1}},\frac{2\alpha p}{\sqrt{2\alpha p-2}} \Bigg) ,
\end{equation}
where $P(\nu,u,x)$ and $Q(\nu,u,x)$ are the associated Legendre
functions of the first and second kind, respectively.

To find the energy spectrum, let us use the annihilation and
creation operators [36] and [37]. By substitution of
\begin{equation}
x=\sqrt{\frac{\hbar}{2m\omega}}(a^{\dag}+a) \quad\quad\quad,\quad\quad\quad p=i\sqrt{\frac{m\hbar \omega}{2}}(a^{\dag}-a) ,
\end{equation}
in Eq. (3) we can write the Hamiltonian of generalized harmonic
oscillator as
\begin{equation}
H=H=\frac{P^{2}}{2m}+\frac{1}{2}m \omega^{2} X^{2} ,
\end{equation}
and then obtain the spectrum of oscillator, $E_{n}=<H>_{n}$, as
\begin{equation}
\epsilon_{n}=2n+1+5\alpha^{2} \gamma^{2}[8\alpha^{2}\gamma^{2}(n+\frac{1}{2})(n^{2}+n+\frac{3}{2})+3(n^{2}+n+\frac{1}{2})] ,
\end{equation}
where
$$
\varepsilon_{n}=\frac{E_{n}}{\frac{\hbar \omega}{2}} \quad\quad\quad,\quad\quad\quad \gamma=\sqrt{\frac{m\hbar \omega}{2}} .
$$

\section{ Generalized Coherent States  }
\quad Within the context of classical mechanics, a physical system
is described by states which are points of its phase space. In
quantum mechanics, the system is described by states which are
vectors in a Hilbert space. There exist superpositions of quantum
states which have many features (properties or dynamical behaviors)
analogous to those of their classical counterparts: they are the
so-called "coherent states". Coherent states were introduced by
Schr\"{o}dinger in 1926 [38] while he was studying the
one-dimensional harmonic oscillator system. These states were
rediscovered by Glauber [39] and Klauder [40] at the beginning of
1960s. The phrase "Coherent States" was proposed by Glauber in the
context of quantum optics. Glauber found them while he was studying
the electromagnetic correlation function. Indeed, these states are
superpositions of Fock states of quantized electromagnetic field
that, up to a complex factor, are not modified by the action of
photon annihilation operators. He also realized that these states
have the interesting property of minimizing the uncertainty
Heisenberg relation. Thus, one could say that these states are the
quantum states with the closest behavior to a classical system.
Coherent states are localized wave packets in position and momentum
spaces and in time are not broadening, in fact remain coherent.
Although there are many ways to construct coherent states, in this
paper we're looking for Klauder's approach and use the version of
the generalized Heisenberg algebra [41]. So, for constructing the
standard coherent states of the harmonic oscillator, we build a
state which is an eigenstate of the annihilation operator of the
generalized Heisenberg algebra. If $|\lambda>$ be a coherent state,
it can be described as an eigenstate of the annihilation operator
\begin{equation}
a|\lambda>=\lambda |\lambda> .
\end{equation}
With expanding $|\lambda>$ in terms of constant states $|n>$, we can
write coherent states in the following form
\begin{equation}
|\lambda>=N(\lambda) \sum _{n=0}^{\infty}\frac{\lambda ^{n}}{N_{n-1}!}|n> ,
\end{equation}
where $N(\lambda)$ is the normalization coefficient, by definition
$N_{n} ! \equiv N_{0}N_{1} \ldots N_{n}$ and by consistency
$N_{-1}!\equiv 1$. It is important to note that Klauder's coherent
states should satisfy the following minimal set of conditions

I) Normalizability
$$
<\lambda|\lambda>=1 ,
$$

II) Continuity in the label
$$
|\lambda-\lambda'|\rightarrow 0 \quad\quad ; \quad\quad \|\: |\lambda>-|\lambda'> \|\rightarrow 0 ,
$$

III) Completeness
$$
\int d^{2} \lambda \, \omega(\lambda) |\lambda><\lambda|=1 .
$$
Since the aforesaid approach implies $N_{n-1}^{2}=\alpha _{n}-\alpha
_{0}$ and $\alpha _{n}=\varepsilon _{n}$, using Eq. (26) we obtain
\begin{equation}
N_{n-1}^{2}=n \Biggl\{ 2+5\alpha ^{2} \gamma ^{2} \Big[ 8\alpha ^{2} \gamma ^{2} (n^{2}+\frac{3}{2}n+2)+3(n+1) \Big]  \Biggr\}.
\end{equation}
To satisfy the normalizability condition and using Eq. (28),
normalization coefficient can be written as
\begin{equation}
N^{2}(|\lambda|)=\frac{1}{\sum _{n=0}^{\infty}\frac{|\lambda|^{2n}}{(N_{n-1}!)^{2}}} .
\end{equation}
Now, we can study the behavior of normalization coefficient by
depicting of $N(|\lambda|)$ for several values of $\beta \equiv
\alpha \gamma$, see Fig. 1. As the figure shows, for $\alpha
\rightarrow 0$ (harmonic oscillator without GUP) $N(|\lambda|)$ goes
to $e^{-\frac{|\lambda|^{2}}{2}}$.

\begin{figure}
\flushleft\leftskip+14em{\includegraphics[width=2.5in]{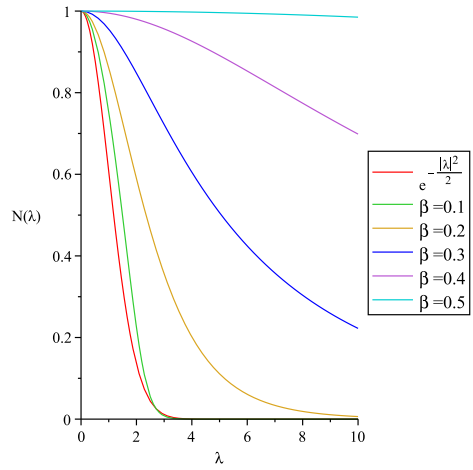}}
\hspace{3cm}\caption{\label{fig:1} Normalization Coefficient for the
GUP-Corrected Harmonic Oscillator.}
\end{figure}

In the absence of GUP and in the ordinary quantum mechanics ($\alpha
\rightarrow 0$), the probability distribution of photons in a
coherent state is given by Poisson distribution
\begin{equation}
P(n,|\lambda|)=|<n|\lambda>|^{2}=\frac{e^{-|\lambda|}}{n!}|\lambda|^{n} .
\end{equation}
Now, in the presence of minimal length and maximal momentum, the
probability is no longer Poissonian, namely
\begin{equation}
P(n,|\lambda|,\alpha)=|<n|\lambda>|^{2}=N^{2}(|\lambda|)\frac{|\lambda|^{2n}}{(N_{n-1}!)^{2}} .
\end{equation}
In Fig. 2, the schematic behavior of the probability distribution
for the GUP-corrected harmonic oscillator is shown for several
values of $\beta$. As the figure shows, for $\alpha \rightarrow 0$
(harmonic oscillator without GUP) P tends to Poisson distribution.

\begin{figure}
\flushleft\leftskip+14em{\includegraphics[width=2.5in]{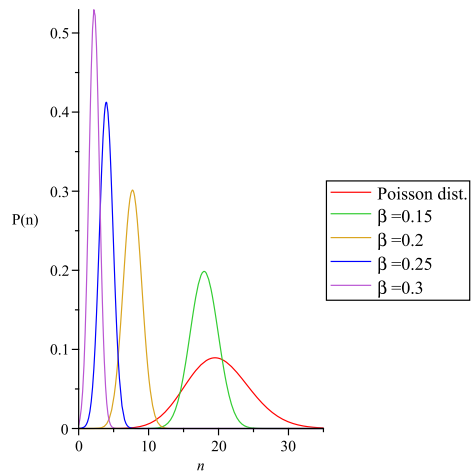}}
\hspace{3cm}\caption{\label{fig:1} The Probability Distribution
for the GUP-Corrected Harmonic Oscillator.}
\end{figure}

\section{Summary and Conclusion}
\quad According to the existence of non-zero uncertainty in the
position and momentum, various physical concepts need to review,
which one of them is the coherence. So, in this paper, we reviewed
the formulation of the generalized uncertainty principle and also
Hilbert space representation of quantum mechanics in the presence of
both minimal observable length and maximal observable momentum.
Then, we have obtained the energy eigenfunctions and spectrum of
energy for a generalized harmonic oscillator in the context of GUP
which implies the both mentioned cutoffs. We showed that because of
the GUP effects, there is a complex mass dependence in energy
spectrum of oscillator. Afterwards, we have studied the coherent
states of the generalized harmonic oscillator. Though, there is no
difference in the definition of coherent states in the GUP
framework, there are some considerable implications due to the
gravitational effects. Therefore, we investigated the general
behavior of normalization coefficient and probability distribution
in terms of GUP parameter.\\

\end{document}